\documentclass[manuscript, nonacm]{acmart}
\AtBeginDocument{%
  }

\begin{document}

\title[Evaluating Explanation Failures as Explainability Pitfalls in Language Learning Systems]{Ceci n'est pas une explication: Evaluating Explanation Failures as Explainability Pitfalls in Language Learning Systems}

%

\author{Ben Knight}
\authornote{Both authors contributed equally}

\affiliation{%
 \institution{Oxford University Press}
 \city{Oxford}
 \country{UK}}
 \email{elt-bench@oup.com}

\author{Wm. Matthew Kennedy}
\authornotemark[1]
\affiliation{%
  \institution{Oxford Internet Institute}
  \institution{University of Oxford}
  \city{Oxford}
  \country{UK}
}

\author{Danielle Carvalho}
\affiliation{%
  \institution{Oxford University Press}
  \city{Oxford}
  \country{UK}}

  \author{Isaac Pattis}
\affiliation{%
  \institution{Oxford University Press}
  \city{Oxford}
  \country{UK}}

  \author{James Edgell}
\affiliation{%
  \institution{Oxford University Press}
  \city{Oxford}
  \country{UK}}

\renewcommand{\shortauthors}{Knight et al.}

\begin{abstract}
AI-powered language learning tools increasingly provide instant, personalised feedback to millions of learners worldwide. However, this feedback can fail in ways that are difficult for learners—and even teachers—to detect, potentially reinforcing misconceptions and eroding learning outcomes over extended use. We present a portion of L2-Bench, a benchmark for evaluating AI systems in language education that includes (but is not limited to) six critical dimensions of effective feedback—diagnostic accuracy, awareness of appropriacy, causes of error, prioritisation, guidance for improvement, and supporting self-regulation. We analyse how AI systems can fail with respect to these dimensions. These failures, which we argue are conducive to "explainability pitfalls", are AI-generated explanations that appear helpful on the surface but are fundamentally flawed, increasing the risk of attainment, human-AI interaction, and socioaffective harms. We discuss how the specific context of language learning amplifies these risks and outline open questions we believe merit more attention when designing evaluation frameworks specifically. Our analysis aims to expand the community's understanding of both the typology of explainability pitfalls and the contextual dynamics in which they may occur in order to encourage AI developers to better design safe, trustworthy, and effective AI explanations.  

\end{abstract}





\maketitle

\thispagestyle{empty}
\renewcommand{\thefootnote}{\fnsymbol{footnote}}
\footnotetext{Accepted to Misleading Impacts Resulting from AI Generated Explanations (MIRAGE) Workshop @ IUI 2026}

\section{Introduction}

Feedback in the form of explanations is a critical part of any learning system. In AI systems design, explanations have been proposed as a key mechanism through which more explainable, transparent, and ultimately accountable AI systems can be developed. If an AI system’s actions can be explained to a user, that user may, in theory, be more empowered to steer the system towards more beneficial outcomes \cite{Teso2023Interactive}. Elsewhere, in educational spaces, explanations provided as feedback in learning contexts supports both knowledge and cognitive process dimensions of individual learning \cite{Anderson2001Taxonomy, Urgo2019Taxonomy}. Clear feedback improves learners’ ability to identify areas of weakness and understand target behaviours \cite{Stobart2019Effective, Hyland1998Feedback, Wiliam2011Embedded}. As it is typically delivered by trusted instructors or peers, it also facilitates social processes that are conducive to knowledge formation \cite{Bandura1977SocialLearning, Brookhart2008Feedback}. In this vein, feedback is more than information transmission; it engages and motivates learners, provided it is constructive, specific, and timely \cite{Vygotsky1978Mind}.    
 
These human-computer interaction and pedagogical considerations intersect in the development of digital learning systems. Digital learning system design is aware of the value and function of feedback in digital pedagogy \cite{VanLehn2011Tutoring, Woolf2010Tutors, Piech2015Modeling}. Some of the first Intelligent Tutoring Systems viewed appropriate feedback as the core function of the system itself \cite{Papert1980Mindstorms}.  Developers of new digital learning systems—those powered by generative AI technologies—maintain this emphasis on feedback as essential to effective AI pedagogy \cite{Jurenka2024ResponsibleEduAI}. Many indeed perceive a new opportunity: recent AI systems may be especially good at delivering constructive, specific, and timely feedback, given their reliance on powerful large language models or multimodal models that perform well on tasks critical to human interaction, such as natural language processing or image recognition \cite{Holmes2024AIED, Woodhead2023Tutors, Pease2023Pedagogical, Jurenka2024ResponsibleEduAI, LearnLM2024Gemini, LearnLM2025GeminiArena}. Although little evidence on the efficacy of such systems exists, many efforts to understand the real-world capabilities of AI learning systems are ongoing \cite{LearnLM2025RCT}.   
 
At the same time that we await more evidence that AI powered learning systems are indeed capable of supporting learning, and, therefore, providing effective feedback in educational contexts, several researchers and educational stakeholders have warned that AI systems may be prone to failures that can have substantial effects on digital learning \cite{HolmesTuomi2022State}. Specifically, although AI tools can generate instant and apparently personalized feedback at scale, the interaction design that structures how feedback is delivered can mask flaws in its quality. These flaws are not the result any intent on the part of system designers to deceive learners (or other educational stakeholders); They occur despite considerable system post-training and fine-tuning for pedagogically appropriate subject matter, interaction design (e.g. optimizing for long conversations), and instructional goals \cite{Kennedy2026Vernacularized}. Given that many of these failures present themselves most clearly in feedback—for example, in conversational turns that follow student completion of a learning task—they can also be exceedingly difficult to detect as those most directly exposed are the least likely to be sufficiently expert to perceive them \cite{HolmesMiao2023UNESCO, Kennedy2026Vernacularized}. Failures can therefore compound over time.  
 
For all these reasons, we believe they may best be approached as “explainability pitfalls” \cite{Ehsan2021Pitfalls}. Problematizing these kinds of failures in AI for education systems in this way has implications both for system designers hoping to improve their systems through monitoring usage data and for learners and instructors who may find that such failures mistakenly reinforce misunderstandings or weaken trust between students and instructors. Understanding when these kinds of failures may occur, with what variety, and to what effect is crucial to advancing any AI for education system design. This need is particularly pronounced in AI for language learning \cite{Stobart2019Effective}.  
 
To this end, our paper discusses our efforts to design L2-Bench \cite{edgell2026accuracyrobustevaluationmethodology}, an evaluation benchmark to assess the efficacy of AI systems in language learning design, which decomposes the tasks that underpin a ‘learning experience designer’ into a hierarchical taxonomy of competencies . In this workshop paper, we focus specifically on the portion of our recently proposed taxonomy of competencies that concern the application of “giving feedback” effectively. In the first section of the paper, for each of these six areas, we briefly outline the core criteria for good feedback in language learning and comment on how AI systems may fail in each dimension. In the second section, we enumerate how AI failures discussed above may produce conditions conducive to explainability pitfalls. We also note how the specific context of language learning can create especially pronounced attainment, human-AI interaction, and other socioaffective risks that are specific to learning contexts and particularly intensified in language-learning contexts \cite{McNamara2023LLMLanguageLearning}. Ultimately, we aim to expand our understanding of both the typology of explainability pitfalls, and the contextual dynamics in which they may occur, lessons we believe can be generally useful to all AI system designers who are interested in safe, trustworthy, and effective AI explanations.

\section{Giving Effective Feedback: a Review of Core Properties} 

Research on effective feedback identifies several key properties that distinguish pedagogically sound feedback from superficial or harmful responses \cite{Stobart2019Effective, Hattie2007Feedback}. Within L2-Bench, we operationalise the competency of "giving feedback" through a structured evaluation framework examining six critical dimensions. These extend beyond simple error correction to encompass the full range of considerations essential to effective language teaching feedback.   

\subsection{Diagnostic Accuracy} 

\textit{Does the AI correctly identify the type and locus of a problem (grammar vs. lexis vs. discourse) and infer intended meaning?  
} 

Effective feedback begins with accurate diagnosis—understanding not just that an error occurred, but precisely where and what kind \cite{Stobart2019Effective}. Hallucinations and overconfidence from AI systems are well-documented and are particularly problematic when the nature of an error is ambiguous. For instance, using the wrong past form of a verb could stem from not knowing verb forms, misunderstanding contextual significance, or overlooking subject-verb agreement. Each diagnostic hypothesis suggests different pedagogical interventions, and misdiagnosis can reinforce rather than correct misconceptions.

\subsection{Awareness of Appropriacy} 

\textit{Does feedback evaluate appropriateness (formality, stance, audience, genre conventions, cultural pragmatics) rather than just correctness?  
} 

Language learning extends beyond grammatical accuracy to encompass using language appropriately for different contexts and purposes \cite{Hyland1998Feedback, HolmesMiao2023UNESCO, LearnLM2025GeminiArena}. This is particularly salient in the communicative approach to language teaching, where priority is placed on conveying meaning in social situations \cite{Stobart2019Effective}. A job application may be grammatically flawless yet employ informal language that creates a negative impression due to its inappropriacy for the professional context. Feedback focusing solely on correctness while ignoring appropriacy fails to prepare learners for authentic communication.

\subsection{Causes of Error} 

\textit{Does the feedback hypothesize why the error occurred (e.g., transfer from L1 word order, article systems, collocation patterns)?  
} 

Understanding the underlying cause—rather than simply identifying the surface manifestation—is a hallmark of expert teaching \cite{Stobart2019Effective}. A teacher experienced with students of a particular mother tongue can predict where negative transfer is likely to occur. AI systems lacking either sufficiently robust internal representations of a given problem space or causal reasoning capabilities may provide feedback that addresses errors superficially without helping learners understand why those errors occurred, limiting transfer of learning to new contexts.

\subsection{Prioritization of Feedback} 

\textit{Does the feedback effectively prioritise what is most useful to the learner?  
}

More feedback is not necessarily better feedback \cite{Stobart2019Effective, Hattie2007Feedback}. When learners are overwhelmed with corrections, they struggle to identify priorities. A vital aspect of effective feedback is identifying which areas will be most useful, drawing upon understanding of what is developmentally appropriate for the learner's proficiency level and current learning goals. AI systems that indiscriminately correct every error risk inducing interaction harms such as anxiety and, paradoxically, reducing effectiveness by failing to prioritise \cite{Kennedy2026Vernacularized}.

\subsection{Guidance for Improvement} 

\textit{Does the feedback help the learner to do better next time?  
} 

The fundamental purpose of feedback is to support future improvement \cite{Hattie2007Feedback, Shute2008Formative}. Effective feedback must strike a delicate balance: offering sufficient information without simply providing answers \cite{Stobart2019Effective}. This aligns with "offering strategies rather than solutions"—an approach that develops learner autonomy rather than dependence \cite{Jurenka2024ResponsibleEduAI, Kennedy2026Vernacularized}. For example, rather than correcting all errors, effective feedback might indicate that a paragraph contains three subject-verb agreement errors and invite the learner to locate and correct them. AI systems providing solutions without strategic guidance (a high likelihood unless some sort of memory is implemented) may short-circuit the learning process \cite{Bastani2024HarmLearning}. Furthermore, an instructor’s own familiarity with AI systems and their outputs (including explanations) may vary. This may subsequently lead them to use AI system explanations differently to how system designers intended, with adverse consequences, as Ehsan et al \cite{Ehsan2024WhoXAI} document in other domains.

\subsection{Supporting Self-regulation} 

\textit{The development of self-regulatory skills is central to long-term learning success \cite{Stobart2019Effective, Wiliam2011Embedded}. Does the AI encourage self-regulated learning (goal setting, monitoring, strategy choice) or foster passive reliance?  
} 

A commonly observed phenomenon with AI tools is that learners become dependent on them without developing their own metacognitive capabilities \cite{Selwyn2019Robots, Gerlich2025Cognitive}. Such dependencies may be intensified by design choices that promote an anthropomorphization of an instructional character \cite{Abercrombie2023Mirages, Kennedy2026Vernacularized}. Feedback must be designed to avoid this dependency trap, ensuring learners do not receive answers without actively engaging with the learning process. This corresponds to the highest level in Absolum's feedback framework—"provocative prompts" that encourage deeper reflection even when success criteria have been met \cite{Stobart2019Effective}. AI systems that fail in this dimension may produce short-term performance gains while undermining lifelong learning capabilities \cite{Nie2025GPTSurprise}.

\section{The Anatomy of an Explainability Pitfall in AI for Language Learning} 

The six dimensions of effective feedback outlined above are not merely pedagogical ideals—they can function as risk models. They can be used to structure efforts to identify failure modes in which AI systems can generate feedback that appears helpful on the surface but is fundamentally flawed; either failing to support learning or actively undermining it. These failures constitute what we might view as "explainability pitfalls" in language learning contexts. Context is a critical factor in producing explainability pitfalls as, in the words of Weidinger et al \cite{Weidinger2023Sociotechnical}, context “co-determines” the harms and benefits AI systems (or, for that matter, any sociotechnical system) produces. This is especially true in educational spaces, whose norms and values differ in important ways from those operating in society in general \cite{Kennedy2024Vernacularizing}.  
 
Drawing on this and our preliminary evaluation of frontier AI models on L2-Bench tasks for the “giving feedback” competency, we observe several characteristic failure patterns that remain difficult for learners (and often teachers) to detect.

\subsection{Over-Correction} 

If the AI system corrects every minor error, learners may become anxious or focus excessively on form rather than meaning, reducing fluency \cite{Stobart2019Effective, Hattie2007Feedback}. This failure in prioritisation manifests as an explainability pitfall when feedback overwhelms rather than guides, leaving learners unable to distinguish critical issues from minor ones.  

\subsection{Ambiguous or Unhelpful Feedback} 

Feedback like "Incorrect" without explanation doesn't teach. Learners need actionable guidance (e.g., why it's wrong and what to do next) \cite{Stobart2019Effective, Shute2008Formative}. AI systems that provide corrected versions without explaining reasoning foster dependency and short-circuit learning—resolving the immediate problem without building capacity to solve similar problems independently.  

\subsection{Ignoring Communicative Competence} 

Systems that only target grammar accuracy may neglect pragmatic and discourse-level skills, leading to unnatural language use \cite{Hyland1998Feedback}. This represents a fundamental explainability failure: the AI cannot adequately explain ‘why’ certain language choices are problematic because it lacks robust models of social context and communicative purpose. Although some systems are improving in-context learning capabilities via expanded context windows, negative verification tasks remain challenging to most LLMs \cite{Fu2025AbsenceBench}.  

\subsection{Cultural and Contextual Misalignment} 

AI systems trained primarily on certain varieties of English may provide feedback reflecting hidden biases about "correct" usage \cite{Klenowski2009Assessment} or may cluster around a central tendency, producing a homogenizing effect \cite{HolmesMiao2023UNESCO, Agarwal2024Homogenize}. Feedback failing to account for learners' cultural backgrounds or learning goals represents an explainability failure at the level of contextual assumptions—the system cannot adequately explain ‘for whom’ and ‘in what contexts’ its guidance is intended \cite{HolmesMiao2023UNESCO}, which can result in irrelevant or even harmful feedback.  

\subsection{Bias and Inconsistency} 

Poorly designed systems might give inconsistent feedback or reinforce incorrect patterns in uncertainty, harming learning progress. When AI systems provide definitive diagnoses without expressing appropriate uncertainty—even when multiple interpretations are plausible—misdiagnosis can reinforce rather than correct misconceptions. In part, this results from methods used to train AI models–methods that OpenAI researchers argue “reward guessing over acknowledging uncertainty” \cite{Kalai2025Hallucinate}. This is also a data preparation problem–a common step in preparing either training data or data for evaluation datasets is to remove samples with ambiguous properties.

\section{Future work} 
By participating in this workshop, we aim to engage participants to aid our continued exploration of concrete examples of explainability pitfalls in AI-generated feedback for language learning. We are particularly interested in understanding where the general failure modes described in section one and their specific presentation in language learning contexts described in section two can usefully generalize to broader research on explainability pitfalls in general. From our point of view, the development of L2-Bench's feedback evaluation framework raises three critical questions that we welcome perspectives on from the broader community working on AI explanations and educational systems that may inform AI explanations (and risks entailed in generating them) more broadly:  

\subsection{Cross-Cultural and Contextual Validity}

Language learning feedback practices vary significantly across cultural and educational contexts. How can AI systems be evaluated for their ability to adapt feedback styles appropriately to diverse learner populations, particularly in global English language teaching contexts where learners' first languages, educational backgrounds, and cultural expectations of teacher-student interaction differ substantially? Some preliminary investigations from the field of adversarial evaluation exist \cite{Kundu2025RedTeaming}, but more work clearly needs to be done here. 

\subsection{Explainability in Uncertain Pedagogical Situations}

Many feedback scenarios in language learning lack clear "correct" answers—for instance, when multiple formulations are acceptable, when appropriacy depends on unstated contextual factors \cite{Gabriel2024EthicsAssistants}, or when diagnostic uncertainty is high. How should AI systems communicate uncertainty in such cases, and what evaluation criteria can distinguish pedagogically appropriate expressions of uncertainty from evasiveness or overconfidence?  

\subsection{Multi-turn Interaction }

A major limitation of our current evaluation benchmark is that it does not yet account for multi-turn interaction. Although some harms do present themselves within the “foundational unit” of conversation—single turn adjacency pairs \cite{Schegloff1973Closings}—many do not \cite{Graesser1995Dialogue}. Indeed this is the key mechanic driving low detection rates of compounding errors in accuracy but also broader sociocultural and contextual misalignment. Yet no AI for education development or evaluation team appears to have made much progress on this front as accounting for multi-turn interaction is an essential condition for evaluating pedagogical efficacy as well as socioaffective risk \cite{LearnLM2025GeminiArena, Kennedy2026Vernacularized}.

\section{Conclusion} 

The unique role that feedback plays in language learning—extending beyond binary correctness judgments to encompass appropriacy, learner development, and socioaffective dimensions—makes it both a critical and challenging domain for AI evaluation. Assessing feedback failures therefore requires a different approach than simply computing an accuracy metric. Instead, we find utility in approaching these types of failures as inadequacies in human-AI interaction design for educational contexts. We first review core principles of feedback as practiced in language learning contexts and discuss where we hypothesize AI systems may perform inadequately. We then discuss how such failures may produce explainability pitfalls and document initial efforts to formalize a portion of a new AI for language-learning evaluation benchmark (L2-Bench) drawing upon that concept. We conclude by observing that several critical challenges remain both for AI for education system designers and evaluation developers. We seek the advice of the broader XAI community about how best to make progress on these important questions, and, in so doing, advance the responsible development of AI systems that genuinely support language learning and boost learning outcomes.

\bibliographystyle{ACM-Reference-Format}
\bibliography{mirage_bib}

\end{document}